\documentclass[12pt]{iopart}

\usepackage{graphicx}
\begin{document}

\title{Fabrication of FeSe$_{1-x}$ superconducting films with bulk properties}

\author{Soon-Gil Jung$^1$, N H Lee$^1$, Eun-Mi Choi$^1$, W N Kang$^{1}$, Sung-Ik Lee$^2$, Tae-Jong Hwang$^3$ and D H Kim$^3$ }
\address{$^1$BK21 Physics Division and Department of Physics, Sungkyunkwan University, Suwon 440-746, Republic of Korea}
\address{$^2$Department of Physics, National Creative Research Initiative Center for Superconductivity, Sogang University, Seoul 121-742, Republic of Korea}
\address{$^3$Department of Physics, Yeungnam University, Gyeongsan 712-749, Republic of Korea}
\ead{wnkang@skku.edu}

\begin{abstract}
We have fabricated high-quality FeSe$_{1-x}$ superconducting films
with a bulk $T_c$ of $11 - 12$ K on different substrates,
Al$_2$O$_3$(0001), SrTiO$_3$(100), MgO(100), and LaAlO$_3$(100), by
using a pulsed laser deposition technique. All the films were grown
at a high substrate temperature of 610 $^\circ$C, and were
preferentially oriented along the (101) direction, the latter being
to be a key to fabricating of FeSe$_{1-x}$ superconducting thin
films with high $T_c$. According to the energy dispersive
spectroscopy data, the Fe:Se composition ratio was 1:0.90$\pm$0.02.
The FeSe$_{1-x}$ film grown on a SrTiO$_3$ substrate showed the best
quality with a high upper critical magnetic field [$H_{c2}$(0)] of
56 T.
\end{abstract}

\pacs{74.70.Xa, 74.78.-w, 74.25.F-}

\maketitle

\section{Introduction}
A new iron age has occurred in the field of superconductivity and
started when the discovery of the LaFeAsO$_{1-x}$F$_{x}$
superconductor in 2008 gave rise to a huge interest in depressed
superconductivity division \cite{1}. Over the past two years, an
enormous number of papers have been published and more than 70 new
iron-based superconductors have been discovered \cite{2}. Generally,
iron-based superconductors are classified as REFeAsO (1111 family,
RE: rare earths), AEFe$_2$As$_2$ (122 family, AE: alkaline earths),
AFeAs (111 family, A: Li, Na), and FeCh (11 family, Ch: chalcogens).
Among these families, a tetragonal phase FeSe is an attractive
material for finding the origin of iron-based superconductors due to
its simple crystal structure \cite{3}. It has just superconducting
layers without block layers, and its superconducting critical
temperature ($T_c$) is very sensitive to change in the strain.
Although the $T_c$ of bulk FeSe is as low as 8.5 K, its $T_c$ can be
raised to 37 K under pressure \cite{4, 5}. In addition, the $T_c$ of
FeSe$_{0.5}$Te$_{0.5}$ thin films can be enhanced to 21 K by
controlling the biaxial strain \cite{6}.

However, compared with Te- or S-doped superconductors \cite{7, 8},
research on pure FeSe superconductors is rarely carried out.
Synthesis of single-phase FeSe superconductors with a tetragonal
structure is not easy because it has several types of phases, such
as tetragonal (PbO-type), hexagonal (NiAs-type), and Fe$_7$Se$_8$
phases \cite{3, 9}. Moreover, most of them, except for the
tetragonal phase, are known to be ferromagnetic phases \cite{10}.
According to recent reports \cite{11}, the possibility of
ferromagnetism was presented in FeSe thin films with a tetragonal
structure at temperature near room temperature, but this issue is
still under debate.

Nevertheless, the observation of superconductivity in the tetragonal
phase of FeSe by Hsu $et~al$ \cite{3} has offered a new member (11
family) to the class of iron-based superconductors. Also, an
expectation of a higher $T_c$ in FeTe than in FeSe, which is caused
by strong paring \cite{12}, has brought about active research on the
Fe-Te-Se system \cite{6, 7, 13}. If the intrinsic properties are to
be investigated precisely, undoped FeSe or FeTe single crystals with
large sizes or thin films are required because isovalent
substitution induces lattice strain due to the different ionic
sizes.

Very recently, Han $et~al$ \cite{14} and Si $et~al$ \cite{15}
reported the appearance of superconductivity in FeTe films due
tensile stress effects and oxygen incorporation, respectively, while
no superconductivity was observed in bulk samples \cite{8}. In the
case of FeSe films, only a few papers have been published \cite{16,
17, 18}, and the experimental results are still unsatisfactory
compared to those for bulk samples. Furthermore, there has been
little progress in the growth of pure FeSe films whereas some
research groups have fabricated high-quality Te-doped FeSe thin
films \cite{6, 19}.

In this work, we report the fabrication of FeSe$_{1-x}$ films with a
bulk $T_c$ of $11 - 12$ K on several types of substrates,
Al$_2$O$_3$(0001) [AO], SrTiO$_3$(100) [STO], MgO(100), and
LaAlO$_3$(100) [LAO], by using a pulsed laser deposition (PLD)
technique. All samples were grown at a high substrate temperature of
610 $^\circ$C and showed high upper critical fields with a (101)
preferred orientation.

\section{Experiments}
Homemade FeSe targets having a stoichiometric ratio of Fe:Se = 1:1
were used in this study, and the laser beam was generated by using a
Lambda Physik KrF excimer laser ($\lambda$ = 248 nm). Our PLD system
is described in detail elsewhere \cite{20, 21}. A laser energy
density of 1.15 J/cm$^2$, a high repetition rate of 48 Hz, and high
substrate temperature ($T_s$) of 610 $^\circ$C were used for the
deposition of FeSe$_{1-x}$ films in a high vacuum state of $\sim$
10$^{-5}$ Torr. The high frequency of the laser is favorable for the
fabrication of films containing highly-volatile materials \cite{20,
21}. High growth rates of $5 - 6$ nm/s were obtained under these
growth conditions. The compositional ratio of the deposited films
was Fe:Se = 1:0.90$\pm$0.02.

The crystal phases and orientations of the fabricated FeSe$_{1-x}$
films were investigated by using x-ray diffraction (XRD). The
surface morphology and thickness were checked by using scanning
electron microscopy (SEM). The elemental composition ratio was
analyzed by using energy dispersive spectroscopy (EDS) based on the
results for a standard stoichiometric sample. A vibrating sample
measurement system (SQUID-VSM, Quantum Design) and a physical
property measurement system (PPMS, Quantum Design) were used for the
study of their superconducting properties.

\section{Results and discussion}

The temperature dependences of the resistivity ($\rho$) of the
FeSe$_{1-x}$ films are shown in figure 1. A Se deficiency of $x$ =
0.10$\pm$0.02, which was comparable with previously reported samples
\cite{3, 18}, was examined by using EDS. All of the deposited films
showed a $T_{c,90\%}$ of $11 - 12$ K and a $T_{c,10\%}$ of $\sim 8$
K, and their transition widths ($\Delta$$T_c$) decreased with
increasing residual resistivity ratio (RRR); these data are
summarized in table 1. All these data are very close to the previous
results for bulk samples \cite{3, 4, 22} and show a higher
$T_{c,10\%}$ and a narrower $\Delta$$T_c$ than previously reported
\cite{16, 17, 18}. In figure 1(a), the $\rho$($T$) of the film grown
on a STO substrate is multiplied by 5 because it has small values
compared with those of the other film. The change of slope around
120 K seems to originate from the influence of a spin-density wave
(SDW) \cite{12, 22, 23}.

\begin{figure}[t!]
\begin{center}
\includegraphics[scale=0.4, angle=0]{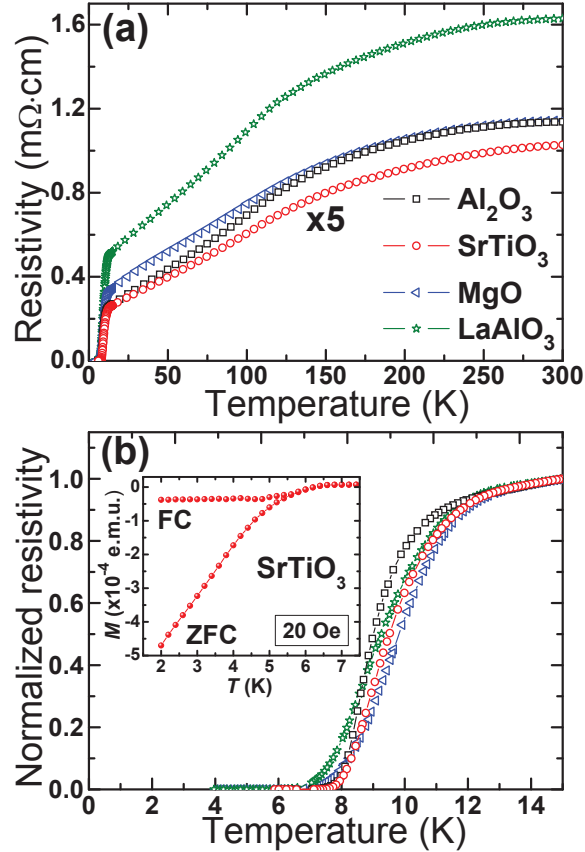}
\caption{(a) Resistivity as functions of temperatures for
FeSe$_{1-x}$ films grown on several types of substrates. and (b) an
enlarged view near the superconducting transition temperature, with
the resistivity values normalized to those at 15 K. The inset of (b)
shows the field-cooled (FC) and the zero-field-cooled (ZFC)
measurements for an FeSe$_{1-x}$ film grown on a SrTiO$_{3}$
substrate at $H$ = 20 Oe.} \label{FIG.1}
\end{center}
\end{figure}

Figure 1(b) is a magnified view near the superconducting transition
temperature; the resistivity values are normalized to those at 15 K.
The inset shows the temperature dependence of the magnetization
($M$) of the FeSe$_{1-x}$ film fabricated on a STO substrate. The
clear diamagnetic signal indicates that it has good bulk
superconductivity. Although the irreversible temperature of $\sim$ 6
K in the $M - T$ curve is lower than the $T_c$ obtained from
$\rho$($T$), obtaining the superconducting signal from magnetization
measurements in FeSe superconductors is not easy due to its
intrinsic magnetism \cite{24} or its magnetic impurities, such as
hexagonal FeSe phases and Fe impurities \cite{3, 25}.

\begin{table} [b!]
\caption{\label{tabone}Summary of the $\rho$($T$) data of figure 1:
superconducting transition temperatures $T_{c,90\%}$ (90\% of
$\rho_n$) and $T_{c,10\%}$ (10\% of $\rho_n$), where $\rho_n$ is the
normal state resistivity near $T_c$, $\Delta T_c$ (=
$T_{c,90\%}-T_{c,10\%}$), and RRR (= $\rho_{300K}/\rho_n$). Fe:Se
data are the composition ratios, obtained from EDS, of the
FeSe$_{1-x}$ films fabricated on the different substrates.}

\begin{indented}
\lineup
\item[]\begin{tabular}{@{}*{6}{l}}
\br
\0\0substrate&$T_{c,90\%}$(K)&$T_{c,10\%}$(K)&\m$\Delta$$T_c$(K)&RRR&Fe:Se\cr
\mr \0\0Al$_2$O$_3$&10.5&8.2&2.3&4.5&1:0.91\cr
\0\0SrTiO$_3$&11.4&8.3&3.1&4.2&1:0.92\cr
\0\0MgO&11.7&8.1&3.6&3.4&1:0.88\cr
\0\0LaAlO$_3$&11.2&7.7&3.5&3.3&1:0.91\cr \br
\end{tabular}
\end{indented}
\end{table}

\begin{figure}
\begin{center}
\includegraphics[scale=0.35, angle=90]{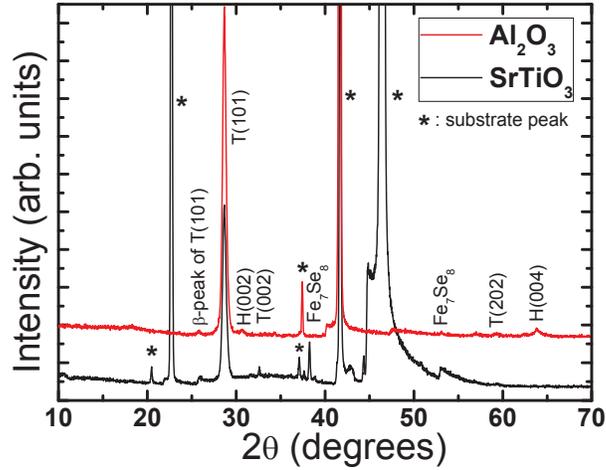}
\caption{$\theta-2\theta$ scan XRD patterns of the FeSe$_{1-x}$
films grown on Al$_{2}$O$_{3}$(0001) and SrTiO$_{3}$(100)
substrates. The preferred orientation is seen to be along the (101)
direction (T: tetragonal, H: hexagonal).} \label{FIG.2}
\end{center}
\end{figure}

\begin{figure}
\begin{center}
\includegraphics[scale=0.35, angle=90]{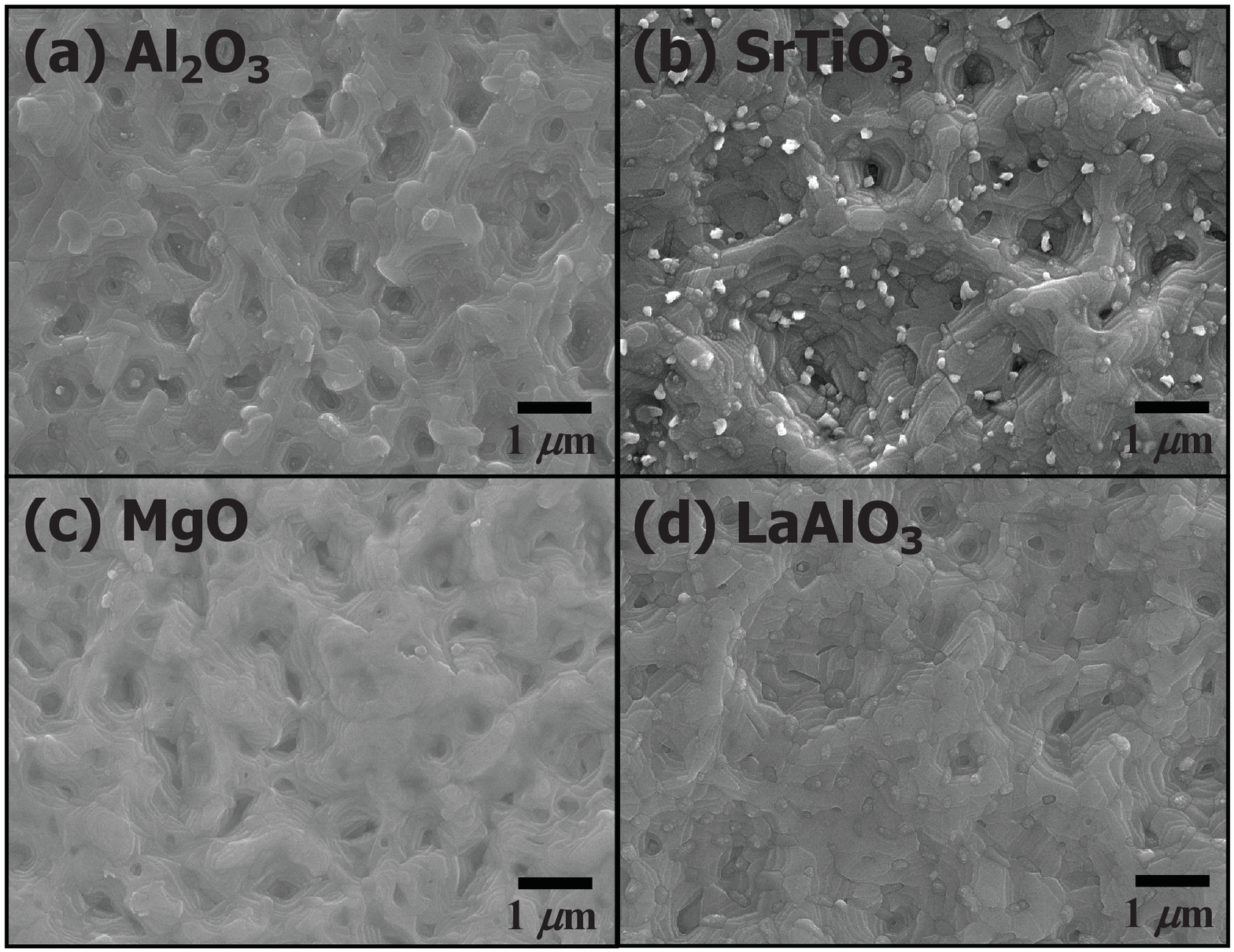}
\caption{SEM images of the FeSe$_{1-x}$ films grown on several
different substrates: (a) Al$_{2}$O$_{3}$, (b) SrTiO$_{3}$, (c) MgO,
and (d) LaAlO$_{3}$. All surface morphologies show well-linked
states without cracks.} \label{FIG.3}
\end{center}
\end{figure}

Figure 2 shows the $\theta - 2\theta$ scan XRD patterns of
FeSe$_{1-x}$ films prepared on the Al$_2$O$_3$(0001) and
SrTiO$_3$(100) substrates. Our samples are mainly oriented along the
(101) direction of the tetragonal (T) structure, but other phases
exist in tiny amounts, such as hexagonal (H) FeSe and Fe$_7$Se$_8$.
The lattice constants for the $a$- and $c$-axis are 3.769 {\AA} and
5.490 {\AA}, respectively, which are close to the values observed in
the bulk samples \cite{3}. No peak shift is detected regardless of
the substrate. We believe that this is due to the growth direction
of the film being along the (101) direction rather than along the
$c$-axis or the ${ab}$-axis. Wang $et~al$ \cite{16} reported that
the orientation of the T(101) direction is the key to fabricating of
FeSe$_{1-x}$ films with strong superconductivity.

The surface morphologies of the FeSe$_{1-x}$ films on the AO, STO,
MgO, and LAO substrates were checked by using scanning electron
microscopy (SEM), as shown in figure 3. The SEM images, (a), (b),
(c), and (d), show that compared with previous reports \cite{18,
26}, all films have well-connected morphologies without cracks,
which is one factor in obtaining higher-quality samples with a
sharper superconducting transition. The white tiny grains in figure
3(b) were shown to be Fe particles by using EDS. Films grown on LAO
had larger Fe lumps than ones grown on STO. These Fe lumps, which
were ferromagnetic, are a possible source of the proximity effect
that accompanies a suppression of $T_c$ \cite{27}. These locally
weakened superconducting regions generate a broad superconducting
transition. The thicknesses determined from the SEM cross-sectional
images in (a), (b), (c), and (d) are 1.5, 1.6, 1.6, and 1.8 $\mu$m,
respectively.

\begin{figure}
\begin{center}
\includegraphics[scale=0.4, angle=90]{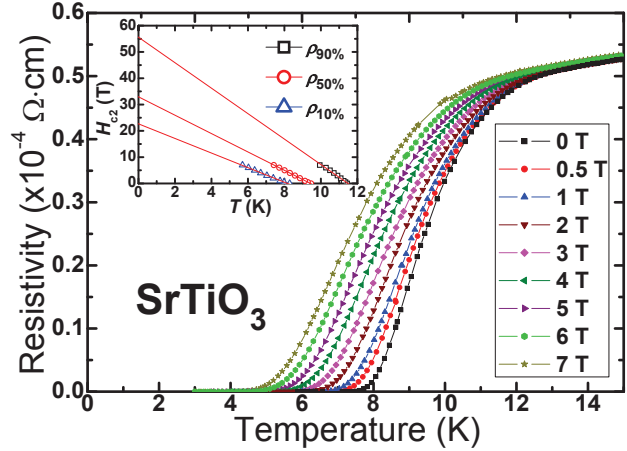}
\caption{$\rho$($T$) curves for FeSe$_{1-x}$ films on SrTiO$_{3}$
substrates under magnetic fields up to 7 T for a field parallel to
the substrate plane. The inset presents the linear fittings for
$H_{c2}$($T$) at $\rho_{10\%}$, $\rho_{50\%}$, and $\rho_{90\%}$.}
\label{FIG.4}
\end{center}
\end{figure}

Finally, we investigated the temperature dependence of the upper
critical magnetic field ($H_{c2}$($T$)) for the FeSe$_{1-x}$ films.
Figure 4 displays the $\rho - T$ curves of the film prepared on a
STO substrate for various magnetic fields from 0 to 7 T, and the
inset shows the fitting of $H_{c2}$($T$). The data points were
obtained from the $\rho - T$ curves for three defined criteria;
$\rho_{90\%}$ (90\% of $\rho_n$), $\rho_{50\%}$ (50\% of $\rho_n$),
and $\rho_{10\%}$ (10\% of $\rho_n$) are equivalent to critical
temperatures of $T_{c,90\%}$, $T_{c,50\%}$, and $T_{c,10\%}$,
respectively. The values of $H_{c2}$(0) at $\rho_{90\%}$ are 58, 56,
50, and 57 T for films grown on AO, STO, MgO, and LAO substrates,
respectively, which were estimated by using a linear extrapolation
\cite{4}. These high values of $H_{c2}$(0) open up the possibility
for practical applications, provided we can grow better films with
higher values of $T_c$.

\section{Conclusions}
In summary, we have deposited FeSe$_{1-x}$ films with bulk
superconducting properties on various types of substrates,
Al$_2$O$_3$(0001), SrTiO$_3$(100), MgO(100), and LaAlO$_3$(100), by
using a pulsed laser deposition system. All films were
preferentially oriented along the (101) direction, and showed upper
critical fields as high as 50 T. Our results showed that
(101)-oriented FeSe$_{1-x}$ films with bulk superconductivity could
be fabricated at high growth temperature regardless of type of
substrate.
\\

\ack {This work was supported by the Korea Science and Engineering
Foundation (KOSEF) grant funded by the Korea Government (Grant No.
MEST, No. R01-2008-000-20586-0) and by the Postdoctoral Research
Program of Sungkyunkwan University (2009)}


\section*{References}
\begin{thebibliography}{99}

\bibitem{1}
Kamihara Y, Watanabe T, Hirano M and Hosono H 2008 {\it J. Am. Chem.
Soc.} {\bf 130} 3296

\bibitem{2}
Zhi-An Ren and Zhong-Xian Zhao 2009 {\it Adv. Mater.} {\bf 21} 1

\bibitem{3}
Hsu F C {\it et al} 2008 {\it Proc. Natl. Acad. Sci. USA} {\bf 105}
14262

\bibitem{4}
Mizuguchi Y,Tomioka F, Tsuda S, Yamaguchi T and Takano Y {\it Appl.
Phys. Lett.} 2008 {\bf 93} 152505

\bibitem{5}
Medvedev S {\it et al} 2009 {\it Nature Mater.} {\bf 8} 630

\bibitem{6}
Bellingeri E {\it et al} 2010 {\it Appl. Phys. Lett.} {\bf 96}
102512

\bibitem{7}
Sales B C, Sefat A S, McGuire M A, Jin R Y, Mandrus D and
Mozharivskyj Y 2009 {\it Phys. Rev. B} {\bf 79} 094521

\bibitem{8}
Mizuguchi Y, Tomioka F, Tsuda S, Yamaguchi T and Takano Y 2009 {\it
J. Phys. Soc. Jpn.} {\bf 78} 074712

\bibitem{9}
Schuster W, Mikler H and Komarek K L 1979 {\it Monats. Chem.} {\bf
110} 1153

\bibitem{10}
Hirone T and Chiba S 1956 {\it J. Phys. Soc. Jpn.} {\bf 11} 666

\bibitem{11}
Liu K W, Zhang J Y, Shen D Z, Shan C X, Li B H, Lu Y M and Fan X W
2007 {\it Appl. Phys. Lett.} {\bf 90} 262503

\bibitem{12}
Subedi A, Zhang L, Singh D J and Du M H 2008 {\it Phys. Rev. B} {\bf
78} 134514

\bibitem{13}
Taen T, Tsuchiya Y, Nakajima Y and Tamegai T 2009 {\it Phys. Rev. B}
{\bf 80} 092502

\bibitem{14}
Han Y, Li W Y, Cao L X, Wang X Y, Xu B, Zhao B R, Guo Y Q and Yang J
L 2010 {\it Phys. Rev. Lett.} {\bf 104} 017003

\bibitem{15}
Si W, Jie Q, Wu L, Zhou J, Gu G, Johnson P D and Li Q 2010 {\it
Phys. Rev. B} {\bf 81} 092506

\bibitem{16}
Wang M J {\it et al} 2009 {\it Phys. Rev. Lett.} {\bf 103} 117002

\bibitem{17}
Nie Y F, Brahimi E, Budnick J I, Hines W A, Jain M and Wells B O
2009 {\it Appl. Phys. Lett.} {\bf 94} 242505

\bibitem{18}
Han Y, Li W Y, Cao L X, Zhang S, Xu B and Zhao B R 2009 {\it J.
Phys.: Condens. Matter} {\bf 21} 235702

\bibitem{19}
Si W, Lin Zhi-Wei, Jie Q, Yin Wei-Guo, Zhou J, Gu G, Johnson P D and
Li Q 2009 {\it Appl. Phys. Lett.} {\bf 95} 052504

\bibitem{20}
Choi Eun-Mi, Jung Soon-Gil, Lee N H, Kwon Young-Seung, Kang W N, Kim
D H, Jung Myung-Hwa, Lee Sung-Ik and Sun L 2009 {\it Appl. Phys.
Lett.} {\bf 95} 062507

\bibitem{21}
Jung Soon-Gil, Lee N H, Seong W K, Kang W N, Choi Eun-Mi and Lee
Sung-Ik 2008 {\it Supercond. Sci. Technol.} {\bf 21} 085017

\bibitem{22}
Zhang S B {\it et al} 2009 {\it Supercond. Sci. Technol.} {\bf 22}
075016

\bibitem{23}
Imai T, Ahilan K, Ning F L, McQueen T M and Cava R F 2009 {\it Phys.
Rev. Lett.} {\bf 102} 177005

\bibitem{24}
Lee K -W, Pardo V and Pickett W E 2008 {\it Phys. Rev. B} {\bf 78}
174502

\bibitem{25}
Patel U {\it et al} 2009 {\it Appl. Phys. Lett.} {\bf 94} 082508

\bibitem{26}
Backen E, Haindl S, Niemeier T, Huhne R, Freudenberg T, Werner J,
Behr G, Schultz L and Holzapfel B 2008 {\it Supercond. Sci.
Technol.} {\bf 21} 122001

\bibitem{27}
Buzdin A I 2005 {\it Rev. Mod. Phys.} {\bf 77} 935

\end{thebibliography}
\end{document}